\documentclass[10pt,journal,compsoc]{IEEEtran}

\ifCLASSOPTIONcompsoc
  \usepackage[nocompress]{cite}
\else
  \usepackage{cite}
\fi

\usepackage{cite}
\usepackage{amsmath,amssymb,amsfonts}
\usepackage{algorithm}
\usepackage{algorithmic}
\usepackage{arevmath}     
\usepackage{graphicx}
\usepackage{multirow}
\usepackage{tabularx}
\usepackage{textcomp}
\usepackage{xcolor}

\begin{document}

\title{Deep-AIR: A Hybrid CNN-LSTM Framework for Fine-Grained Air Pollution Forecast}

\author{Qi Zhang\textsuperscript{*}, 
        Victor OK Li, ~\IEEEmembership{Fellow,~IEEE},
        Jacqueline CK Lam\textsuperscript{*}, ~\IEEEmembership{Member,~IEEE},
        Yang Han
\IEEEcompsocitemizethanks{\IEEEcompsocthanksitem Corresponding authors. The authors are with the Department of Electrical and Electronic Engineering, The University of Hong Kong, Pok Fu Lam, Hong Kong.\protect\\
E-mail: \{zhangqi, vli, jcklam, yhan\}@eee.hku.hk}

}

\IEEEtitleabstractindextext{%
\begin{abstract}
Poor air quality has become an increasingly critical challenge for many metropolitan cities, which carries many catastrophic physical and mental consequences on human health and quality of life. However, accurately monitoring and forecasting air quality remains a highly challenging endeavour. Limited by geographically sparse data, traditional statistical models and newly emerging data-driven methods of air quality forecasting mainly focused on the temporal correlation between the historical temporal datasets of air pollutants. However, in reality, both distribution and dispersion of air pollutants are highly location-dependant. In this paper, we propose a novel hybrid deep learning model that combines Convolutional Neural Networks (CNN) and Long Short Term Memory (LSTM) together to forecast air quality at high-resolution. Our model can utilize the spatial correlation characteristic of our air pollutant datasets to achieve higher forecasting accuracy than existing deep learning models of air pollution forecast.
\end{abstract}

}

\maketitle

\IEEEraisesectionheading{\section{Introduction}\label{sec:introduction}}

\IEEEPARstart{A}{s} the world's greatest developing country, China has experienced rapid economic development over the last three decades. Such development has brought about rapid deterioration in air quality, resulting in catastrophic human health consequences including respiratory and cardiovascular diseases. Hence, accurately monitoring and forecasting the concentration of PM\textsubscript{2.5} (particulates smaller than 2.5 micrometers in diameter) and other harming pollutants will positively impact human health and dictate the quality of life of our citizens, especially the future generations.

\subsection{Related Work}
Previous air pollution-related studies focused on urban air pollution forecasting via different modelling methodologies. Generally, we can divide the methods of air quality forecasting into two major modelling approaches: the physical-based modelling approach and the data-driven modelling approach.

Physical-based models are based on computational fluid dynamics and chemical reactions. Research was conducted based on pollutant emissions such as fuel consumption, transportation, etc.\cite{arya1999air,pai1997simulation,baklanov2008towards,ji2015environmental}. Air pollution forecast was conducted based on certain theoretical hypotheses. These models are computationally complex and difficult to be applied in a large geographical scale, such as the entire city.

Data-driven models avoided complicated physical models and theoretical hypotheses. They inferred the future air pollution based on patterns learned from existing data. Traditional statistical inference models applied in air quality forecast included linear regression\cite{li2011study}, support vector regression\cite{nieto2013svm}, and autoregressive integrated moving average\cite{kumar2010arima}. \cite{zhu2017extended} proposed a Granger Causality model to identify spatial-temporal causalities among air pollutants, which outperformed other baseline causality learning models. As a newly emerging technique, deep learning approach made full use of the large amount of historical data and outperforms traditional statistical inference methods \cite{zhang2017prediction,ong2016dynamically,li2017long,li2017deep}. \cite{ong2016dynamically} proposed a recurrent neural network for air pollution forecasting in Tokyo, with pre-trained auto-encoder to pre-process the input data. \cite{li2017deep} proposed an LSTM neural network that used urban proxy data to facilitate air quality forecast. \cite{li2017long} extended the LSTM model by combining historical air pollution data with auxiliary data where input data were used to predict air quality in Beijing, and achieved state-of-the-art single-model performance. However, existing data-driven air quality forecasting methods mainly took into account the temporal correlation of the time series data, with limited forecasting accuracy given the loss of spatial information.

\begin{figure}[htbp]
\centerline{\includegraphics[width=\linewidth]{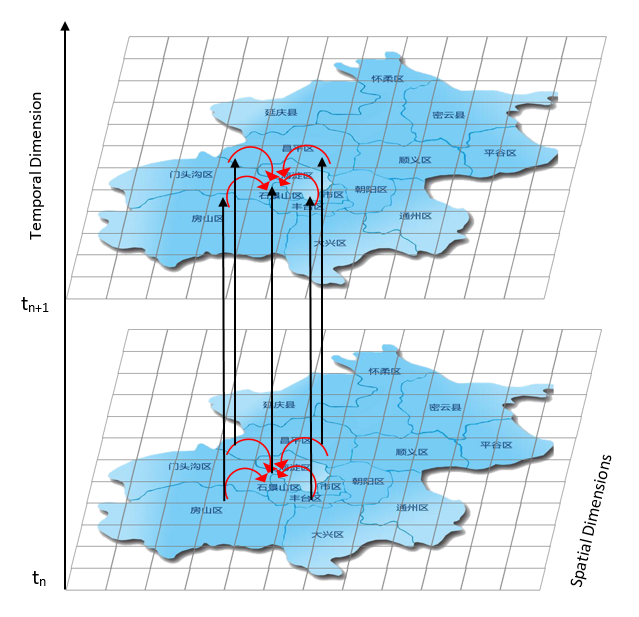}}
\caption{Temporal and spatial correlation of urban dynamics data}
\label{fig:map}
\end{figure}

Apart from air pollution forecast, another challenging research topic is the geographically fine-grained estimation of air pollution. The models mentioned above were only trained for the points (grids) where air pollution data had been provided by monitoring stations, and they were not able to forecast on the points (grids) where air pollution data were unavailable, that is, where monitoring stations were unavailable. To address this challenge, one might use portable ancillary sensors. In \cite{feng2018air}, the author used an ELM-based neural network to process the data generated by wireless sensor networks (WSNs). \cite{yang2018aqnet} deployed programmable on-ground sensors as well as unmanned-aerial-vehicle (UAV) to perform 3D air quality monitoring. These methods had achieved some satisfactory results for geographically fine-grained air pollution estimation, but infrastructure deployment could be highly costly. Zheng et al.\cite{zheng2015forecasting} used a hybrid model with separate temporal predictor and spatial predictor to provide a relatively fine-grained forecast of air quality, but prediction still remained too low a resolution. A more complex deep learning framework that provides high-resolution forecast of air pollution at low deployment cost is much needed for highly populated and highly polluted juristictions, such as China and India.

\subsection{Research Significance}
Air pollution, together with other urban dynamics data, are both temporally and spatially correlated. The correlation of different dimensions is shown in Fig \ref{fig:map}. Firstly, the air pollution at a specific area has been correlated with its historical value, for air quality changes continuously in the real world. Secondly, due to the dispersion of air pollutants, air pollution at a particular location is highly dependent on the air quality of its neighborhood locations. In addition, the meteorology information of any nearby places has to be considered as the dispersion process is significantly affected by wind speed, temperature, pressure, etc.

As mentioned in the previous section, existing research mainly took into account the temporal correlation of air quality data. In some related work\cite{ong2016dynamically, zheng2015forecasting, fan2017spatiotemporal, yi2018deep}, the authors attempted to incorporate the spatial information in air pollution forecasting. However, they simply added the data of nearest stations as the input of their model, without considering the relative location of the data. 

In this paper, we design Deep-AIR, a hybrid deep learning framework for air pollution forecast. It incorporates a convolutional neural network (CNN) component to extract the spatial feature of the data, and a recurrent neural network (RNN) component to learn the temporal feature. The contributions of our work include:
\begin{itemize}
  \item This is the first deep learning model that includes CNN network in learning the spatial feature of air pollution data. With the specially designed structure, we can handle the spatial effect of different urban dynamics of air pollution.
  \item Our framework is the first deep learning model that can generate a geographically fine-grained air pollution forecast without the help of any ancillary sensors.
  \item With the incorporation of spatial features, our proposed hybrid CNN+LSTM model achieves a better forecasting performance than the existing baseline models including ARIMA and pure LSTM.
\end{itemize}

\section{Methodology}
\begin{figure}[htbp]
\centerline{\includegraphics[width=\linewidth]{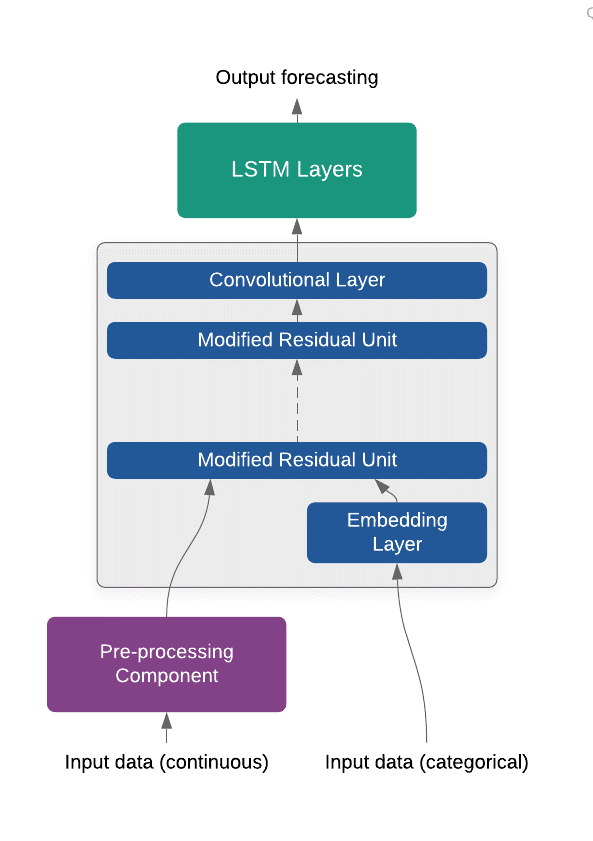}}
\caption{Overview of Deep-AIR, our novel hybrid CNN-LSTM deep learning framework}
\label{fig:structure}
\end{figure}

We design Deep-AIR, a deep learning framework based on the unique spatio-temporal characteristics of urban dynamics data. Our deep learning structure consists of three components. A data pre-processing component uses interpolation methods to deal with the missing data. A residual convolutional neural network component (AirRes) is then used for spatial feature extraction. The top structure consists of two Long Short Term Memory (LSTM) layers, which are used to learn the temporal features of historical air pollution data and generate forecasting results. The overall structure of Deep-AIR is shown in Fig. \ref{fig:structure}

\subsection{Data Pre-processing Component}
We divide the city of Beijing into 3km\(\times\)3km grids. Each grid has its own air pollution data, plus other urban dynamics data as proxy input data. Hence, for each time slot, the input data of the whole city can be seen as an n-channel image. Each pixel corresponds to a grid of the map, and each channel corresponds to one kind of air pollution or other proxy data. Since deep learning models require fixed size of input data, the data pre-processing component is necessary to solve two major challenges brought by the special characteristics of urban dynamics data: the incompleteness of historical records and the geographical sparsity of air pollution monitoring stations.

One commonly used method to overcome missing values is to directly omit the missing values, filling the missing data with zeros or randomly generated values, another method is to fill in the missing values with the latest valid observation or linear interpolation. The first approach creates a huge noise to the dataset since the missing part of the dataset contains necessary information in air pollution forecast. The second method alleviates the loss of information to a certain extent. However, simple linear interpolation can not be used to generate a geographically fine-grained distribution of urban dynamics data since the observed points are sparse.

\begin{table*}[ht]
\caption{Pearson's Coefficients of Urban Dynamics across All Stations}
\begin{center}
\begin{tabular}{|l|l|l|l|l|l|l|l|l|}
\hline
\multirow{2}{*}{Data Type} & \multicolumn{5}{c|}{AQI}                                             & \multicolumn{3}{c|}{Traffic}  \\ \cline{2-9} 
                      & PM\textsubscript{2.5}    & PM\textsubscript{10}        & NO\textsubscript{2}            & CO            & O\textsubscript{3}         & Status    & Speed   & Count   \\ \hline
Value                 & 0.84     & 0.76        & 0.67           & 0.70          & 0.88       & 0.31      & 0.43    & 0.24    \\ \hline
\multirow{2}{*}{Data Type} & \multicolumn{8}{c|}{Meteorology}                                                                     \\ \cline{2-9} 
                      & Pressure & Temperature & Wind direction & Precipitation & Wind speed & \multicolumn{3}{c|}{Humidity} \\ \hline
Value                 & 0.99     & 0.98        & 0.91           & 0.25          & 0.56       & \multicolumn{3}{c|}{0.19}     \\ \hline
\end{tabular}
\label{table:tab1}
\end{center}
\end{table*}

Our model incorporates a two-step pre-processing component to handle data incompleteness in temporal and spatial dimensions separately. We first conduct a linear interpolation of historical data for a short period in the temporal dimension. A Kriging interpolation is then applied to generate a grid dataset for the whole city. To verify the validation of the spatial interpolation, we analyze the spatial correlation of different urban dynamics data between the different stations using Pearson's correlation coefficient:

\begin{equation}
r_{xy}=\frac{{}\sum_{i=1}^{n} (x_i - \overline{x})(y_i - \overline{y})}
{\sqrt{\sum_{i=1}^{n} (x_i - \overline{x})^2(y_i - \overline{y})^2}}
\end{equation}

The results are shown in Table \ref{table:tab1}. All air pollution data and part of the meteorology data of different stations are strongly correlated. Following the correlation test, we perform Kriging interpolation for the dynamics that have a strong correlation (\(R>0.6\)) across stations, and fill in the missing values of other dynamics with zeros. The strong spatial correlation of air pollution concentration of different points also supports the use of the same model for the forecasts of all grids on the map, instead of training a different model for each grid.

\subsection{Deep Residual Component}
Convolutional neural network (CNN) has been widely acknowledged as the best tool for spatial feature extraction, and has been utilized in state-of-the-art Computer Vision studies. However, CNN has hardly been employed in the field of air pollution forecast. \cite{huang2018deep} applied CNN in air pollution forecast, but only one-dimensional convolutional layers were used, and were only applied on the temporal dimension. \cite{liu2018third} used a CNN component in its air pollution forecast framework, but the component was used to process photos. In our framework, after the interpolation component, we obtain a city-wide "picture" of urban dynamics. Hence our model is capable of utilizing the powerful characteristics of CNN in extracting spatial information.

\begin{figure}[htbp]
\centerline{\includegraphics[width=\linewidth]{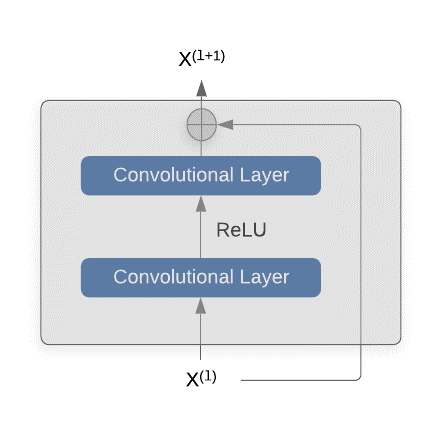}}
\caption{Residual unit}
\label{fig:residual_unit}
\end{figure}

\begin{figure}[htbp]
\centerline{\includegraphics[width=\linewidth]{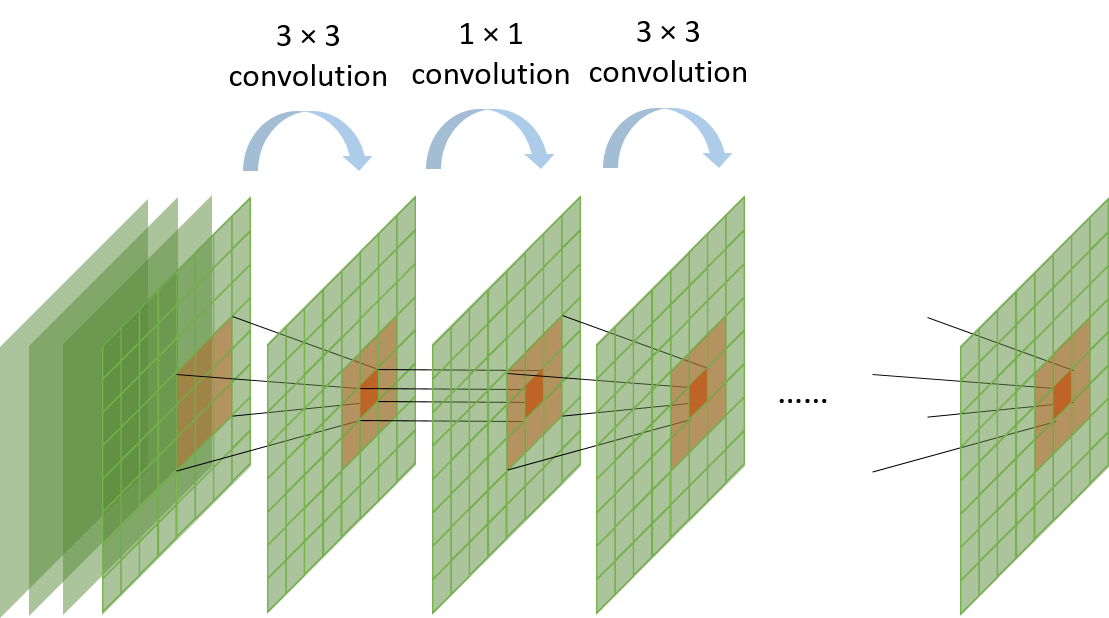}}
\caption{The inner structure of Deep Residual Component}
\label{fig:CNN}
\end{figure}

\textbf{Deep Residual Network} In order to let the model learn as large a scope of information as possible, a CNN with deep structure is needed. For this purpose, we employ Deep residual network (ResNet) in our framework. ResNet\cite{he2016deep} is a type of CNN that has very deep structure due to its internal residual units. A typical residual unit consists of a few convolutional layers and an identity mapping, as shown below:
\begin{equation}
X^{(l+1)}=X^{(l)} + \mathcal{F}(X^{(l)})
\end{equation}
where \(X^{(l)}\) and \(X^{(l+1)}\) denote the input and output matrix of the \(l^{th}\) unit. \(\mathcal{F}\) denotes the mapping function conducted by the convolutional layers. The residual units create shortcut for the information flow, and benefit the training process of very deep networks\cite{he2016deep}. It is also used in other scenarios of urban computing such as traffic flow prediction\cite{zhang2017deep}.

\begin{figure*}[htp]
\centerline{\includegraphics[width=0.75\paperwidth]{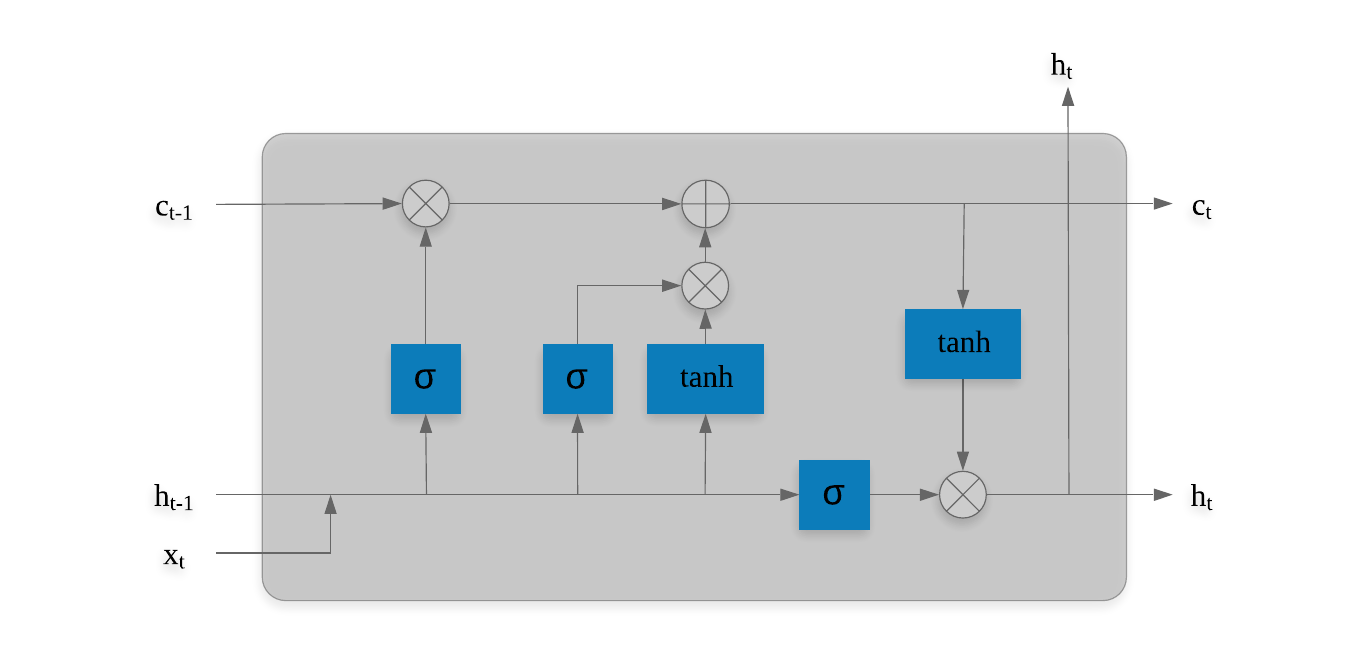}}
\caption{Structure of an LSTM block}
\label{fig:lstm}
\end{figure*}

\textbf{1\(\times\)1 convolution}
Although the input data of our model is "picture" like, the structure of ResNet in our framework need to be modified to adapt to the special characteristics of urban dynamics. As mentioned in the first section, the dispersion of air pollutants is strongly dependent on the meteorology status such as pressure and wind speed of nearby grids. Therefore, we need to facilitate the information exchange of different channels of input data. To address this challenge, we propose the AirRes structure, as shown in Fig. \ref{fig:CNN}. In this structure, 1\(\times\)1 convolutional layers are inserted between the 3\(\times\)3 convolutional layers of residual units. 1\(\times\)1 convolution is widely known for its capability to reduce the number of channels in GoogLeNet architecture\cite{szegedy2015going}. However, it can also facilitate information interflow across channels\cite{lin2013network} because the output of a 1\(\times\)1 convolutional layer is equivalent to a linear combination of different feature maps. Results in Table \ref{table:tab3} show that the 1\(\times\)1 convolution layer improves the performance of ResNet in air pollution forecasting.

To further facilitate the forecast accuracy, we add external auxiliary input data apart from urban dynamics including day of the week and hour of the day. Since these data are of categorical value, they are first provided in the form of one-hot vector and then sent to an embedding layer to be transformed into real-valued vectors. The embedding layer before the AirRes unit can reduce the dimension of the auxiliary data, and hence reduce the parameters and enhance learning efficiency.

\subsection{LSTM Layers}
After the Deep Residual Component has extracted high-level spatial features of each time step, the feature matrix of each time step is sent to a Long Short Term Memory (LSTM) structure as input. LSTM neural network is a special kind of Recurrent Neural Network (RNN) characterized by more complex memory blocks than simple neurons at each time step. A memory block consists of several gates to control the information flow of the internal memory cells. The structure is shown in Fig \ref{fig:lstm}. Due to the carefully designed gates, LSTM networks avoid the problem of gradient vanish/explosion that often occurs in RNN. It can remember long term temporal features and works well on time series data. 

The LSTM layers in our framework take the feature matrix generated by the Deep Residual Component, and output a vector for each grid as the predicted air pollution value.

\section{Experimental Setting and Results}
\subsection{Experiment Setting}

\begin{table}[hb]
\centering
\caption{Urban Dynamics Data Collected in Beijing from Jan, 2017 to Jul, 2018}
\begin{tabular}{|l|l|l|l|}
\hline
& \textbf{Air Quality}    & \textbf{Meteorology} & \textbf{Traffic} \\ \hline
Available Points           & 35     & 18          & 227     \\ \hline
Data Type               & 5      & 6           & 3       \\ \hline
Update Frequency           & 1 hour & 3 minutes      & 10 minutes  \\ \hline
\end{tabular}
\label{table:tab2}
\end{table}

We test the accuracy of the fine-grained air pollution forecast on the dataset we collect in Beijing for 19 months (Jan, 2017- Jul, 2018). Three kinds of urban dynamics data are used as input of the model, as shown in Table \ref{table:tab2}. The air pollution data are collected from 35 public air quality monitoring stations, and the meteorology data are collected from 18 meteorology monitoring stations in Beijing. The traffic data are collected via the web API provided by Gaode Map, and traffic information of 227 major roads of Beijing are used as input. For the data updated more frequently than once per hour, we average them for each hour so that the time-stamps of every kind of data are aligned to 1 hour. We use 80\% of data as training set, 10\% as validation set, and the last 10\% as testing set. The three different sets of data are divided in chronological order.

We divide the city of Beijing into 3km\(\times\)3km grids, so that the training and testing data of each time step is a 50\(\times\)55 map with 16 channels (14 urban dynamics channels and two auxiliary channels). The historical data of all of the stations except for a randomly picked one are used as input. The output is the concentration value of five different air pollutants of that particular station in the coming hour. The output is compared to the ground truth air quality data provided by the chosen monitoring station. We use mean average percentage error (MAPE) as metric to measure the forecasting error as follows:

\begin{algorithm}[t]
\caption{Patch Training for Deep-AIR}
\begin{algorithmic}[1]
\REQUIRE urban dynamics map \(\mathbb{D} = \{\mathcal{D}^{T}\}\), air pollution data map \(\mathbb{Q} = \{\mathcal{Q}^{T}\}\), temporal range \(T\), temporal window length \(W\), station location range \(L\), network structure \(\mathit{f}\), network parameters \(\theta\)

\STATE Temporal interpolation for \(\mathbb{D}\) and \(\mathbb{Q}\)
\REPEAT
    \FOR{\(t\) = 1,2...\(T\)}
        \STATE Obtain urban dynamics map \(\{\mathcal{D}^{t'}\}^{t-W<t'<t}\)
        \STATE Sample \(l\) from \(\{1,2...L\}\)
        \FOR{\(t'\in(t-W,t)\)} 
            \STATE Crop the input patch \(d_{l}^{t'}\) from \(\mathcal{D}^{t'}\) 
            \STATE Spatial interpolation for \(d_{l}^{t'}\) without data of grid \(l\)
        \ENDFOR
        \STATE Obtain air pollution forecast for grid \(l\) at time \(t\): \(y_{l}^{t}=\mathit{f}_{\theta}(d_{l}^{t-W+1},...,d_{l}^{t-1}) \)
        \STATE Collect the ground-truth air pollution value \(q_{l}^{t}\) from \(\mathbb{Q}\)
        \STATE Calculate the loss \(\mathcal{L}=\|q_{l}^{t}-y_{l}^{t}\|^{2}\)
        \STATE BP: \(\theta \gets \theta - \lambda\partial\mathcal{L}/\partial\theta\)
    \ENDFOR
\UNTIL{stopping criteria is met}
\end{algorithmic}
\label{algorithm:alg1}
\end{algorithm}

\begin{equation}
MAPE=\frac{|y_{i} - y_{i}^{*}|}
{y_{i}} \times 100\%
\end{equation}


The missing values of all time series are filled by linear interpolation in the temporal dimension. For the grids without urban dynamics data, the data are filled by Krigging interpolation in spatial dimension. The two channels of categorical auxiliary channels are first processed by an embedding layer, and then sent to the AirRes component together with the urban dynamics channels. The effective scope of the AirRes component is 15\(\times\)15, and we use a patch training algorithm to train the model, as illustrated in Algorithm \ref{algorithm:alg1}. For each iteration, a 15 \(\times\) 15 patch of the map with the selected station at the center is used as the input of the proposed Deep-AIR model, and the output of the model is the air pollution concentration value of the center grid. As the patch move through the whole map, the model can generate a fine-grained forecasting of the whole city. 

The AirRes component consists of 4 residual units, and each residual unit has two 3\(\times\)3 convolution layers with batch normalization and ReLU activation function. 1\(\times\)1 convolution layer is added between each two of the Residual units. For the LSTM component, the hidden size is set to 128, and the window length is set to 48. We use stochastic gradient decent to train the model, and the training process is stopped when the validation error is not improved in the latest 5 epochs.

\subsection{Results}

\begin{figure}[htbp]
\centerline{\includegraphics[width=\linewidth]{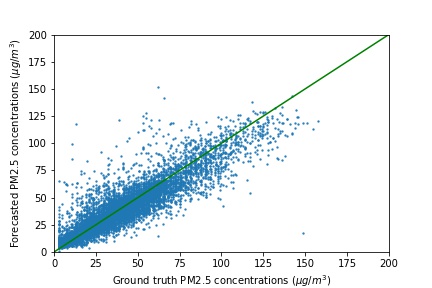}}
\caption{Forecast and ground truth values of PM\textsubscript{2.5} on the test set}
\label{fig:test}
\end{figure}

After choosing the best hyper-parameters based on the performance on the validation set, a test set is utilized to evaluate the forecast error of the model. Fig. \ref{fig:test} is a scatter diagram of the forecast values and ground truth values of PM\textsubscript{2.5} concentration. The scatters are distributed around the line of the identity mapping, and the \(R^{2}\) value between the forecast values and ground truth values is 0.82. If we evaluate the forecast values using the air quality levels according to the ambient air quality standards (GB 2095-2012) released by the Chinese government, the accuracy of the prediction is 80.0\%. It shows that the forecast values are generally consistent with the ground truth values.

\begin{table}[hb]
\centering
\caption{Error Rates of Different Models in Percentage}
\begin{tabular}{|l|l|l|}
\hline
\textbf{Model} & \textbf{Training Error} & \textbf{Testing Error} \\ \hline
LSTM           & 20.1                    & 35.6                   \\ \hline
ConvLSTM       & 18.9                    & 32.7                   \\ \hline
ResNet-LSTM    & 17.5                    & 28.5                   \\ \hline
\textbf{Deep-AIR}  & \textbf{15.7}  & \textbf{27.1}         \\ \hline
\end{tabular}
\label{table:tab3}
\end{table}

Table \ref{table:tab3} compares the average error rate of our model and baseline models. Our proposed Deep-AIR model has the best performance in providing a fine-grained forecast of air pollution across the city. With the AirRes componet, our model can learn the spatial information and achieves better result than a LSTM model. We also compare our model with ConvLSTM, which is a typical structure for spatio-temporal data processing that combines CNN with LSTM. Experiment results show that our model with separate components for spatial and temporal correlations performs better. In addition, compared to a ResNet component without 1\(\times\)1 convolution, our AirRes structure achieves higher forecasting accuracy.

\begin{table}[t]
\centering
\caption{Error Rate of Separate Pollutants}
\begin{tabular}{|l|l|l|l|l|l|}
\hline
Pollutant      & PM\textsubscript{2.5} & PM\textsubscript{10} & NO\textsubscript{2} & CO & O\textsubscript{3} \\ \hline
Training Error & 17.0      & 17.7      & 16.5     & 14.3    & 18.9     \\ \hline
Testing Error  & 31.9      & 25.5      & 26.2     & 16.9    & 37.2    \\ \hline
\end{tabular}
\label{table:tab4}
\end{table}

Table \ref{table:tab4} shows the forecast errors of different kinds of air pollutants. It can be seen that the model gives the best performance in forecasting the concentration of CO, while it can not forecast O\textsubscript{3} with a satisfactory accuracy. This means that the variation trend of O\textsubscript{3} concentration is not as disciplinary as that of other pollutants. The huge gap between the forecast error on the training set and testing set implies that irregular sudden changes of O\textsubscript{3} occur from time to time, and the neural network fails to accurately model the pattern.

\section{Conclusions}
In this study, we have proposed a hybrid deep learning structure (Deep-AIR) for fine-grained air quality forecast. This is the first study that utilizes a Convolutional Neural Network to capture the spatial information of air pollution related data. Experimental results have shown that our hybrid deep learning framework outperforms other existing baseline models in generating fine-grained air pollution forecasts. The improvement of air quality forecast accuracy will carry significant positive impacts on public health and environmental policy making. Our novel methodology taking into account of the unique characteristics of spatio-temporal data can potentially contribute to a wide variety of urban applications, such as crowd flows prediction, traffic control, and related socio-economic research topic at the district and the city level of any highly populated and developing countries, such as China.

\ifCLASSOPTIONcompsoc
  \section*{Acknowledgments}
\else
  \section*{Acknowledgment}
\fi

This research is supported in part by the Theme-based Research Scheme of the Research Grants Council of Hong Kong, under Grant No. T41-709/17-N. We would like to acknowledge Beijing Municipal Environment Monitoring Center and National Meteorological Information Center, China for publicizing air quality and meteorological data of Beijing. We also acknowledge the map service of AutoNavi for making the traffic data in Beijing available.

\bibliographystyle{IEEEtran}
\bibliography{ref}

\begin{IEEEbiography}[{\includegraphics[width=1in,height=1.25in,clip,keepaspectratio]{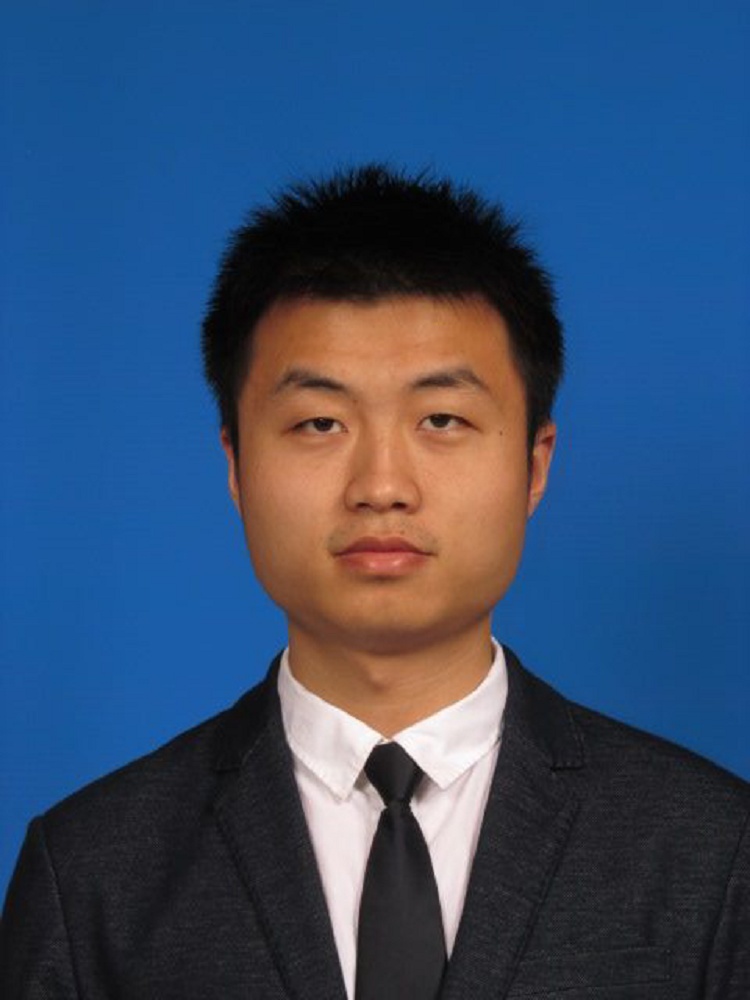}}]{Qi Zhang}
received the BE degree in electronic engineering from Tsinghua University, Beijing, China, in 2017. He is working towards the PhD degree in the Department of Electrical \& Electronic Engineering, The University of Hong Kong (HKU). His research interests include deep learning, spatio-temporal data analytics, and urban computing.
\end{IEEEbiography}

\begin{IEEEbiography}[{\includegraphics[width=1in,height=1.25in,clip,keepaspectratio]{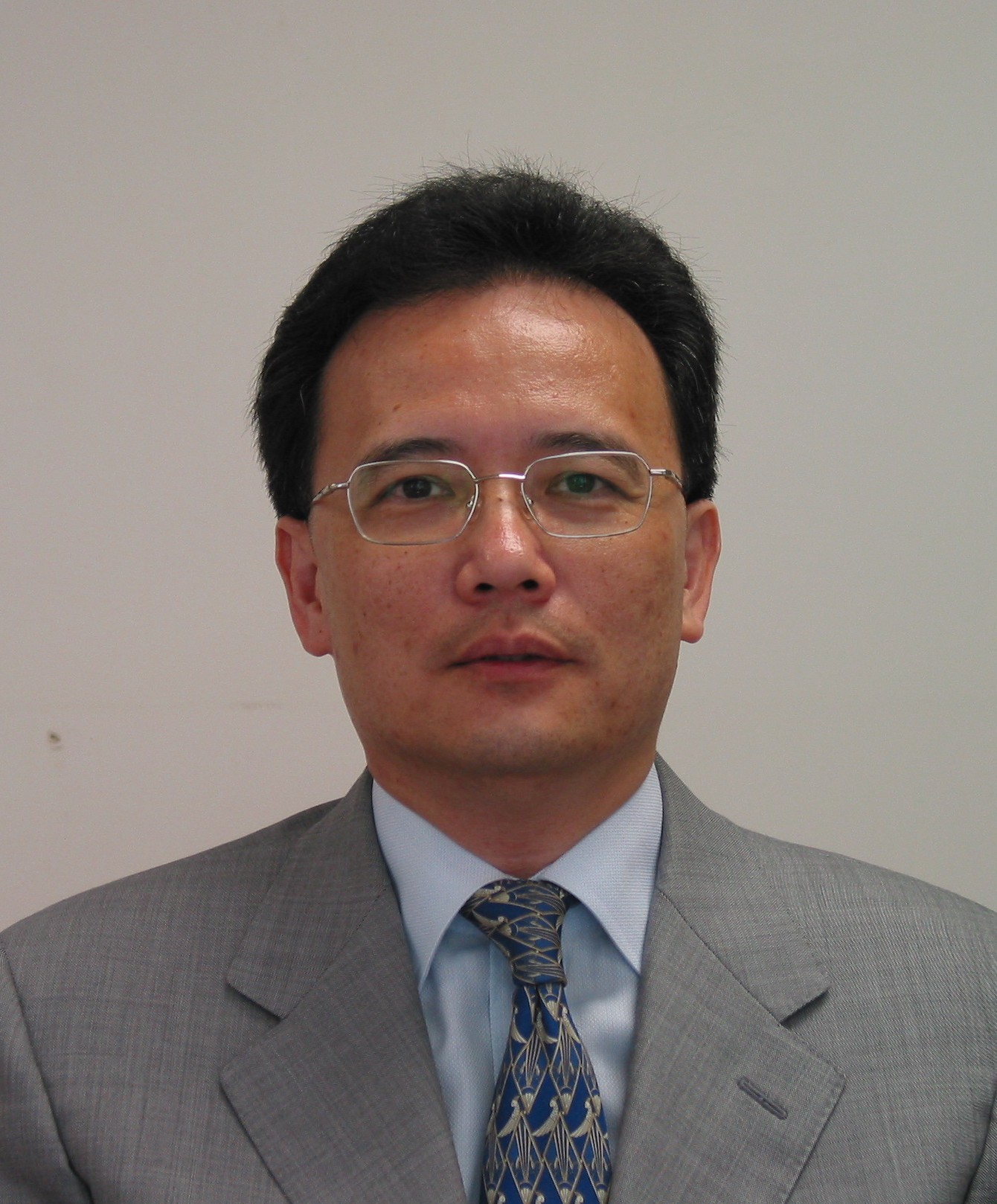}}]{Victor O.K. Li}
Victor O.K. Li  (S’80 – M’81 – F’92) received SB, SM, EE and ScD degrees in Electrical Engineering and Computer Science from MIT.  Prof. Li is Chair of Information Engineering and Cheng Yu-Tung Professor in Sustainable Development at the Department of Electrical \& Electronic Engineering (EEE) at the University of Hong Kong.  He is the Director of the HKU-Cambridge Clean Energy and Environment Research Platform, and of the HKU-Cambridge AI to Advance Well-being and Society Research Platform, which are interdisciplinary collaborations with Cambridge University. He was the Head of EEE, Assoc. Dean (Research) of Engineering and Managing Director of Versitech Ltd.  He serves on the board of Sunevision Holdings Ltd., listed on the Hong Kong Stock Exchange and co-founded Fano Labs Ltd., an artificial intelligence (AI) company with his PhD student.  Previously, he was Professor of Electrical Engineering at the University of Southern California (USC), Los Angeles, California, USA, and Director of the USC Communication Sciences Institute. His research interests include big data, AI, optimization techniques, and interdisciplinary clean energy and environment studies.  In Jan 2018, he was awarded a USD 6.3M RGC Theme-based Research Project to develop deep learning techniques for personalized and smart air pollution monitoring and health management.  Sought by government, industry, and academic organizations, he has lectured and consulted extensively internationally. He has received numerous awards, including the PRC Ministry of Education Changjiang Chair Professorship at Tsinghua University, the UK Royal Academy of Engineering Senior Visiting Fellowship in Communications, the Croucher Foundation Senior Research Fellowship, and the Order of the Bronze Bauhinia Star, Government of the HKSAR.  He is a Fellow of the Hong Kong Academy of Engineering Sciences, the IEEE, the IAE, and the HKIE.
\end{IEEEbiography}

\begin{IEEEbiography}[{\includegraphics[width=1in,height=1.25in,clip,keepaspectratio]{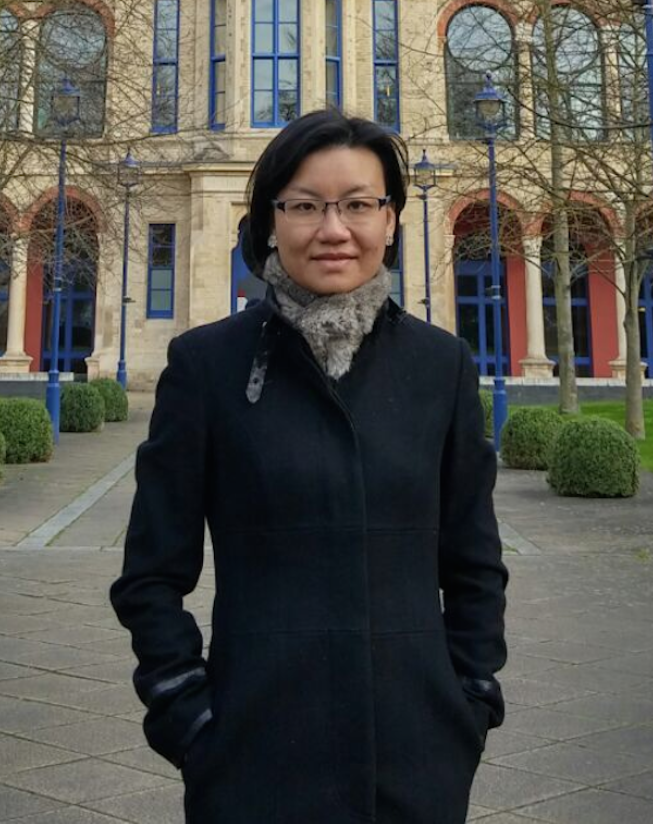}}]{Jacqueline C.K. Lam}
is Associate Professor in the Department of Electrical and Electronic Engineering, the University of Hong Kong and Co-Director of the HKU-Cambridge Clean Energy and Environment Research Platform, and
of the HKU-Cambridge AI to AdvanceWell-being and Society Research Platform. She was the
Hughes Hall Visiting Fellow, before she takes up the Visiting Senior Research Fellow and Associate
Researcher in Energy Policy Research Group, Judge Business School, the University of Cambridge. Her research studies clean energy and environment using interdisciplinary approaches, with a special focus on China and the UK. Her recent research focuses on the use of big data and machine learning techniques to study personalized air pollution monitoring and health management. Her work is published in IEEE, Environment International, Applied Energy, Environmental Science and Policy, and Energy Policy. Jacqueline has received three times the research grants awarded by the Research Grants Council, HKSAR Government, from 2011-2017. The funded amount totalled USD 7.8M in PI or Co-PI capacity. Her recent research study, in joint collaboration with Yang Han and Victor OK Li on PM\textsubscript{2.5} pollution and environmental inequality in Hong Kong, has been published in Environmental Science and Policy, and widely covered by more than 30 local and overseas newspapers and TVs.
\end{IEEEbiography}

\begin{IEEEbiography}[{\includegraphics[width=1in,height=1.25in,clip,keepaspectratio]{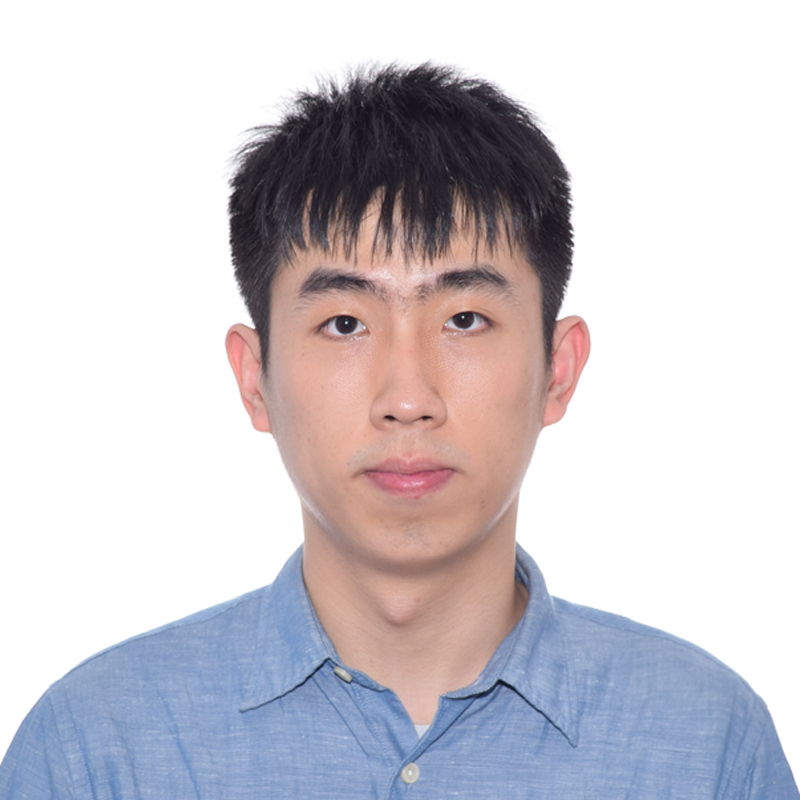}}]{Yang Han}
Yang Han received his MSc degree in Computer Science with distinction from the University of Hong Kong (HKU) in 2014. He obtained his bachelor’s degree from the Department of Information Systems, Beihang University, Beijing, China, in 2013. Currently he is undertaking PhD in the Department of Electrical \& Electronic Engineering, HKU. His recent work focuses on spatio-temporal data analysis and its applications in environmental pollution and policy studies in China. He has published on Environmental Science and Policy.
\end{IEEEbiography}

\end{document}